\documentclass[conference,10pt]{IEEEtran}
\usepackage{epsfig,epsf,rotating,setspace,latexsym,amsmath,amssymb,amsfonts,bm,theorem,subfigure,epstopdf,cite,authblk,bbm,nonfloat,comment}
\usepackage{algorithm}
\usepackage[noend]{algpseudocode}
\usepackage{color}
\usepackage{mathtools}
\usepackage{soul}

\algrenewcommand\algorithmicforall{\textbf{foreach}}
\algrenewcommand\algorithmicindent{.8em}

\newtheorem{theorem}{Theorem}
\newtheorem{lemma}{Lemma}

\newtheorem{remark}{Remark}

\newenvironment{Proof}[1]{\medskip\par\noindent{\bf Proof:\,}\,#1}{{\mbox{\,$\blacksquare$}\par}}

\IEEEoverridecommandlockouts
\allowdisplaybreaks
\begin{document}

\title{Age of Gossip on Generalized Rings} 

\author{Arunabh Srivastava \qquad Sennur Ulukus\\
        \normalsize Department of Electrical and Computer Engineering\\
        \normalsize University of Maryland, College Park, MD 20742\\
        \normalsize  \emph{arunabh@umd.edu} \qquad \emph{ulukus@umd.edu}}

\maketitle

\begin{abstract}
     We consider a gossip network consisting of a source forwarding updates and $n$ nodes placed geometrically in a ring formation. Each node gossips with $f(n)$ nodes on either side, thus communicating with $2f(n)$ nodes in total. $f(n)$ is a sub-linear, non-decreasing and positive function. The source keeps updates of a process, that might be generated or observed, and shares them with the nodes in the ring network. The nodes in the ring network communicate with their neighbors and disseminate these version updates using a push-style gossip strategy. We use the version age metric to quantify the timeliness of information at the nodes. Prior to this work, it was shown that the version age scales as $O(n^{\frac{1}{2}})$ in a ring network, i.e., when $f(n)=1$, and as $O(\log{n})$ in a fully-connected network, i.e., when $2f(n)=n-1$. In this paper, we find an upper bound for the average version age for a set of nodes in such a network in terms of the number of nodes $n$ and the number of gossiped neighbors $2 f(n)$. We show that if $f(n) = \Omega(\frac{n}{\log^2{n}})$, then the version age still scales as $\theta(\log{n})$. We also show that if $f(n)$ is a rational function, then the version age also scales as a rational function. In particular, if $f(n)=n^\alpha$, then version age is $O(n^\frac{1-\alpha}{2})$. Finally, through numerical calculations we verify that, for all practical purposes, if $f(n) = \Omega(n^{0.6})$, the version age scales as $O(\log{n})$.
\end{abstract}

\section{Introduction}
Over the last decade, the number of inter-connected devices has increased rapidly due to the incorporation of wireless capabilities into various devices, as technologies have advanced. This has led to the onset of new applications such as deployment of UAVs and sensors for collecting measurements and surveillance, self-driving car networks, and remote connectivity with home appliances to make life easier. Among such applications, many are time-critical, and it is very important that freshest data is available to carry out the required time-critical tasks. Hence, freshness of information has emerged as an important performance metric in wireless networks. 

It is well-known that latency, an established metric in communication systems, is not sufficient to characterize freshness of information \cite{popovski2022perspective}. In order to better quantify freshness, new metrics have been proposed, such as, age of information \cite{kaul2012real, sun2019age, yatesJSACsurvey}, which has been studied under various settings \cite{yates2018age, banerjee2023re}. Several extended metrics have also been introduced based on real-life inspired applications, including age of incorrect information\cite{maatouk20AOII}, age of synchronization \cite{zhong18AoSync}, binary freshness metric \cite{cho3BinaryFreshness}, and version age of information \cite{yates21gossip, Abolhassani21version, melih2020infocom}.

In this paper, we consider the version age of information metric. The version age of a node in a gossiping network is the number of versions behind the node is when compared to a source node that is generating or observing a random process and has the latest version of the update. \cite{yates21gossip} uses stochastic hybrid systems (SHS) to come up with a set of recursive equations to find the version age of a connected subset of a network. \cite{yates21gossip} also finds that the average version age of a fully-connected network scales as $\theta(\log{n})$ and numerically observes that the version age of a ring network scales as $O(\sqrt{n})$. This work is extended in \cite{buyukates22ClusterGossip} which shows that the version age of a particular arrangement of a network can be improved by having a community structure with smaller networks of the same arrangement. This paper also proves the numerical observation in \cite{yates21gossip} about the version age scaling in a ring. The version age metric is also studied under various different settings. \cite{kaswan22timestomp} studies a network with a timestomping adversary, which can change the timestamps of the updates and fool the nodes into accepting an older version of the update. \cite{kaswan22jamming} studies the metric in the case where there are jamming adversaries. \cite{kaswan23nonpoisson} studies the version age of information in a non-Poisson update setting. \cite{mitra_allerton22} considers version age in an age-sensing multiple access channel. \cite{mitra2023timely} considers opportunistic gossiping protocols that achieve $O(1)$ scaling for version age in distributed multiple access channels. \cite{abd2023distribution} studies the distributions of version age and its moments.

\begin{figure*}[t]
    \centering
    \includegraphics[width = 0.8\linewidth]{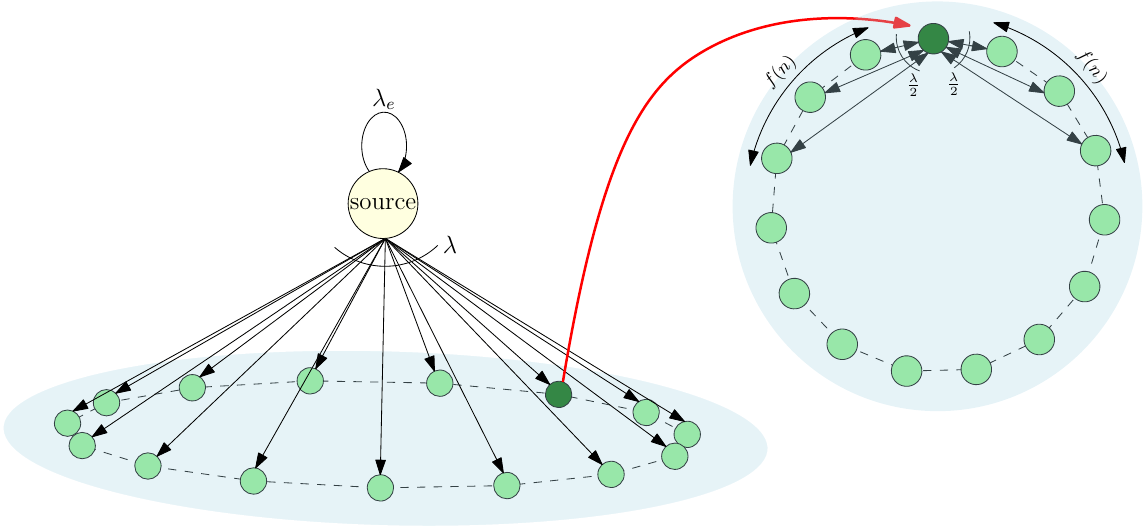}
    \caption{The source node updates itself at rate $\lambda_e$, and disseminates information to nodes arranged in a ring at total rate $\lambda$. Each node communicates with $f(n)$ nodes on each of its sides, thus, communicating with $2f(n)$ nodes in total, where $f(n)$ is a given function in the network size $n$.}
    \label{fig1}
\end{figure*}

In this work, we consider a general network arranged in the form of a ring, see Fig.~\ref{fig1}. The number of neighbors each node has is a positive non-decreasing sub-linear function $f(n)$ of $n$, the number of nodes in the network. Each node gossips with $f(n)$ neighbors on each side, thus, communicating with $2f(n)$ neighbors in total. We find a general upper bound on the version age of a single node in such a network. We recover the results obtained for the fully-connected network and the ring network in \cite{yates21gossip} and \cite{buyukates22ClusterGossip},  by choosing $2f(n)=n-1$ and $f(n)=1$, respectively, in our result. We analyze how the version age varies as the function $f(n)$ grows from $1$ to $n$, and find the upper bound for some special cases, such as functions of the form $f(n) = n^\alpha$ for $0<\alpha<1$.

\section{System Model}
We consider a system where we have a source node generating or observing updates as a rate $\lambda_e$ Poisson process independent of all other processes in the network. The source node disseminates these updates to $n$ nodes in a network. The network of $n$ nodes is denoted by $\mathcal{N}$, and hereafter referred to as the ring network. The $n$ nodes are placed in a ring formation as shown in Fig.~\ref{fig1}. These nodes receive updates from the source node as a combined rate $\lambda$ Poisson process, which can be thought of as a thinned process where each node is being updated by the source node as a rate $\frac{\lambda}{n}$ Poisson process.

Each node in the ring network gossips with its neighbors, which are the nearest $2f(n)$ nodes, i.e., each node gossips with $f(n)$ nodes on each side. Each node sends its version of the update to each neighboring node as a rate $\frac{\lambda}{2f(n)}$ Poisson process, resulting in a total gossiping rate of $\lambda$ per node. 

In order to quantify the freshness of version updates, we use the version age metric. First, we define the counting processes associated with the version updates. Let $N_0(t)$ be a counting process associated with the version updates at the source node, i.e., it increases by $1$ each time the source gets a new version update. In a similar way, we define the version update of node $i$ in the gossip network as the counting process $N_i(t)$, which maintains the latest version at node $i$. Next, we define the version age of node $i$ as $X_i(t) = N_s(t)-N_i(t)$, which quantifies the number of versions node $i$ is behind compared to the source node. We define the version age of the source node to be $0$, since it always has the latest update. Next, we define the version age of a connected subset $S$ of the network as $X_S(t) = \min_{j \in S}X_j(t)$. Finally, we define the limiting average version age of this set $S$ as $v_S = \lim_{t \rightarrow \infty}X_S(t)$.

The evolution of version age of a particular node in the ring network is as follows: If node $i$ in the ring network receives an update directly from the source node, then its version age drops to $0$, since it now has the latest version of the update. If the source node generates or observes a new version of the update, then the version age of node $i$ increases by 1. If a neighboring node shares its version of the update, then node $i$ keeps the new version if it is fresher than the version it has, otherwise it rejects the update and keeps its own version.

We define the rate of information flow from node $i$ to neighboring node $j$ as $\lambda_{ij}$. This is the rate of the Poisson process associated with the updates that node $i$ sends to node $j$. We say that node $i$ is a neighboring node of set $S$ if $\lambda_{ij} > 0$ for some $j \in S$, and define the set of neighboring nodes of $S$ as $N(S)$. Next, we define the rate of information flow from node $i$ into connected set $S$ as $\lambda_i(S) = \sum_{j \in S}\lambda_{ij}$ and $\lambda_i(S) = 0$ if $i \in S$. Similarly, we define the rate of information flow from the source to set $S$ as $\lambda_0(S)$. We call an edge emanating from node $i\notin S$ to node $j\in S$ an incoming edge into $S$ if $\lambda_{ij} > 0$. We call the set of all incoming edges into set $S$ as $E(S)$.

\section{Version Age of a Single Node} \label{sec3}
In this section, we calculate an upper bound for the version age of a single node in the generalized ring network, denoted by $v_1$. We note that the average version age of each node in the network will be the same due to the symmetry in the network. In order to calculate the upper bound, we modify the recursive equations of \cite{yates21gossip}, following the method described in \cite{srivastava2023age}.
\begin{lemma}\label{lemma1}
For any connected subset $S$ of the generalized ring network, we have,
\begin{align}
    v_S \leq \frac{\frac{\lambda_e}{\lambda}+\frac{|E(S)|}{2f(n)}\max_{i \in N(S)}v_{S \cup \{i\}}}{\frac{|S|}{n}+\frac{|E(S)|}{2f(n)}} \label{lemma1-eqn}
\end{align}
\end{lemma}
\begin{Proof}
First, we write the recursive equations from \cite{yates21gossip},
\begin{align}
    v_S = \frac{\lambda_e + \sum_{i \in N(S)}\lambda_i(S)v_{S \cup \{i\}}}{\lambda_0(S) + \sum_{i \in N(S)}\lambda_i(S)} \label{yates-recursion}
\end{align}
In order to find an upper bound, we rearrange (\ref{yates-recursion}) as,
\begin{align}
    \lambda_e = \lambda_0(S)v_S + \sum_{i \in N(S)}\lambda_i(S)\left(v_S - v_{S \cup \{i\}}\right) \label{recur_eq}
\end{align}
Now, we define a function $E_S(i)$ which represents the number of incoming edges to set $S$ that emanate at node $i \in \mathcal{N}$: $E_S(i) = \sum_{k \in S} \mathbb{I}(\lambda_{ik} > 0)$, where $i \in N(S)$ and $\mathbb{I}(\cdot)$ is the indicator function. Then, we partition $N(S)$ into $2f(n)$ sets according to the number of incoming nodes into $S$ from any $i \in N(S)$ as,
\begin{align}
    A_j = \{i \in N(S): E_S(i) = j\}
\end{align}
where $1 \leq j \leq 2f(n)$. Now, we rewrite \eqref{recur_eq} as,
\begin{align}
    \!\!\lambda_e &= \lambda_0(S)v_S + \sum_{j=1}^{2f(n)}\sum_{i \in A_j}\lambda_i(S)\left(v_S - v_{S \cup \{i\}}\right)\\
    &\geq \lambda_0(S)v_S + \sum_{j=1}^{2f(n)}|A_j|\min_{i\in A_j}\lambda_i(S)\left(v_S - v_{S \cup \{i\}}\right)\\
    &\geq \lambda_0(S)v_S + \sum_{j=1}^{2f(n)}|A_j|\min_{i\in A_j}\lambda_i(S)\min_{i\in A_j}\left(v_S - v_{S \cup \{i\}}\right)\!\\
    &= \lambda_0(S)v_S + \sum_{j=1}^{2f(n)}|A_j|\frac{j\lambda}{2f(n)}\left(v_S - \max_{i\in A_j}v_{S \cup \{i\}}\right)\\
    &\geq \lambda_0(S)v_S + \sum_{j=1}^{2f(n)}|A_j|\frac{j\lambda}{2f(n)}\left(v_S - \max_{i\in N(S)}v_{S \cup \{i\}}\right)\\
    &= \lambda_0(S)v_S + |E(S)|\frac{\lambda}{2f(n)}\left(v_S - \max_{i\in N(S)}v_{S \cup \{i\}}\right) \label{last-eqn-lemma1}
\end{align}
where $|E(S)|=\sum_{j=1}^{2f(n)}j|A_j|$ is the total number of incoming edges into set $S$. Rearranging (\ref{last-eqn-lemma1}) together with the substitution $\lambda_0(S) = \frac{\lambda|S|}{n}$ proves the lemma.
\end{Proof}

Next, we develop a further upper bound for (\ref{lemma1-eqn}) in Lemma~\ref{lemma1} by further lower bounding (\ref{last-eqn-lemma1}). For that, we need to identify a lower bound for $|E(S)|$ in (\ref{last-eqn-lemma1}) for a fixed number of nodes in $S$ on a generalized ring. We have the following lemma.

\begin{figure}
    \centering
    \includegraphics[width = \linewidth]{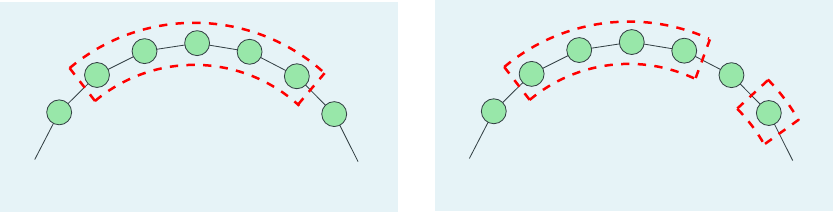}
    \caption{On the left, a contiguous set of $5$ nodes, marked in red, as described in Lemma~\ref{lemma2}. On the right, a set of $5$ nodes that is not contiguous.}
    \label{fig2}
\end{figure}

\begin{lemma}\label{lemma2}
    On a generalized ring, given all connected subsets $S$ such that $|S| = j$, the set which has the minimum number of incoming edges is the contiguous set of $j$ nodes.
\end{lemma}

\begin{Proof}
We denote edges that start at a node in $S$ and end at a node in $S$ as inner edges. We note that each edge can only be an inner edge or an incoming edge. Hence, the sum of the number of incoming edges and twice the number of inner edges is constant and equal to $2jf(n)$. Thus, showing that the set of contiguous nodes has the minimum number of incoming edges is the same as showing that it has the maximum number of inner edges. In this proof, we will show that the number of inner edges is the highest. Fig.~\ref{fig2} shows examples of contiguous and non-contiguous sets.

Let the set of $j$ contiguous nodes be $S_1$, and choose any other connected set of $j$ nodes and call it $S_2$. Next, label each node in both sets as $1_{S_1}, 2_{S_1} \ldots, j_{S_1}$ and $1_{S_2}, 2_{S_2} \ldots, j_{S_2}$, respectively. The labels start at the node at one end of the set and end at the other end, covering each node in order of their position. Now, we compare each $i_{S_1}$ and $i_{S_2}$. We know that both nodes have a total of $2f(n)$ neighbors.

First, we consider the case where $j \leq f(n)$. In this case, $i_{S_1}$ has all the other nodes in $S$ as a neighbor. Hence, it has $j-1$ inner edges associated with it, and this is the highest achievable. $i_{S_2}$ may not have all nodes in $S_2$ as neighbors, and hence has at most the same number of inner edges as $i_{S_1}$. This is true for each consequent node. Hence, adding the number of inner edges of each node in both sets, we see in this case that $S_1$ has more inner edges than $S_2$.

Next, we consider the case where $f(n) < j \leq 2f(n)$. Suppose the number of nodes in the set is $f(n) + k$, then for $i<k$, $i_{S_1}$ shares an inner edge with $f(n)$ nodes on one side and $i-1$ neighbors on the other side, which is the highest possible for its position. If $i>j-k$, $i_{S_1}$ shares an inner edge with $f(n)$ nodes on one side and $n-i$ neighbors on the other side, which is the highest possible for its position. If $k \leq i \leq j-k$, then all nodes in the set share an inner edge with $i_{S_1}$, which again is the highest possible number. Hence, adding the number of inner edges of each node in both sets, we see in this case that $S_1$ has more inner edges than $S_2$.

Next, we consider the case where $2f(n) < j < n-2f(n)$. If $i
\leq f(n)$, then all $f(n)$ neighbors on one side share an inner edge with $i_{S_1}$, and all $i-1$ neighbors on the other side also share an inner edge with $i_{S_1}$, which leads to $i_{S_1}$ having the highest possible number of inner edges as it shares an inner edge with all possible nodes in its position. Hence, $i_{S_2}$ cannot have more inner edges than $i_{S_1}$. Due to symmetry, this is also true for $n-f(n)\leq i$. Also, if $f(n) < i < n-f(n)$, then all $2f(n)$ neighbors of $i_{S_1}$ share an inner edge with it. Once again, $i_{S_2}$ has at most the same number of inner edges. Hence, in this case, adding up the number of inner edges in order for both sets, we conclude that $S_1$ has more inner edges than $S_2$.

Finally, if $j \geq n-2f(n)$, then $\mathcal{N}\backslash S$ has the same number of inner edges as $S$. Hence, following the first and second cases above, the contiguous set has the most inner edges.

From the above four cases, we conclude that the set of contiguous nodes has the highest number of inner edges, and hence, the lowest number of incoming edges.
\end{Proof}

Now, we state and prove our main theorem.

\begin{theorem}
    The version age of a single node in the generalized ring network scales as,
    \begin{align}\label{main_result}
        v_1 = O\left(\log{f(n)} + \frac{\sqrt{n}}{f(n)^{\frac{1}{2}}}\right)
    \end{align}
\end{theorem}

\begin{Proof}
First, using Lemma~\ref{lemma2}, we write the exact lower bounds for incoming edges by counting the number of incoming edges of the sets of contiguous nodes. We have three formulae, corresponding to three regions, as follows:
\begin{enumerate}
    \item $j \!\leq\! f(n)$: \ $|E(S)| \geq 2jf(n) - j(j-1)$
    \item $f(n) \!<\! j \!<\! n\!-\!f(n)$: \ $|E(S)| \geq f(n)(f(n)+1)$
    \item $n\!-\!f(n)\! \leq\! j$: \ $|E(S)| \geq 2(n\!-\!j)f(n) \!-\! (n\!-\!j)(n\!-\!j\!-\!1)$
\end{enumerate}
We obtain the first bound by counting the number of inner edges for each node, which is $j-1$, and then subtracting it from the total number of neighbors $2f(n)$. Then, the number of incoming edges for each node is $2f(n)-j-1$. Since there are $j$ nodes in total, the number of incoming edges into set $S$ is given by $j(2f(n)-(j-1))$. We carry out a similar calculation to count the number of edges in the third case. In the second case, we simply calculate the number of incoming edges. The nearest neighbors on each side has $f(n)$ incoming edges, the second nearest neighbor has $f(n)-1$ incoming edges, and so on. Hence, the total number of incoming edges is given by $2 \times \frac{f(n)(f(n)+1)}{2} = f(n)(f(n) + 1)$.

Next, we calculate the sum of recursive terms for each range. Let the sums of the recursive terms for the ranges be $X$, $Y$ and $Z$, respectively.

\subsection{Range~1}
The upper bound for the recursion for $X$ is,
\begin{align}
    \!\!X &\leq \frac{\lambda_e}{\lambda}\!\!\left(\!\frac{1}{1+\frac{1}{n}}\!\right)\!\!\left(\!1\!+\!\sum_{i=1}^{f(n)}\prod_{j=1}^i \frac{\frac{2jf(n)-j(j-1)}{2f(n)}}{\frac{j+1}{n}+\frac{2(j+1)f(n)-j(j+1)}{2f(n)}}\!\right)\!\!\!\! \label{eq_X_rub}\\
    &\leq \frac{\lambda_e}{\lambda}\left(1+\sum_{i=1}^{f(n)}\prod_{j=1}^i \frac{\frac{2jf(n)-j(j-1)}{2f(n)}}{\frac{2(j+1)f(n)-j(j+1)}{2f(n)}}\right)\\
    &= \frac{\lambda_e}{\lambda}\left(1+\sum_{i=1}^{f(n)}\prod_{j=1}^i \frac{(2f(n)-(j-1))}{2f(n)-j}\frac{j}{j+1}\right)\\
    &\leq \frac{\lambda_e}{\lambda}\left(1+\sum_{i=1}^{f(n)}\frac{1}{i}\frac{2f(n)}{2f(n)-i}\right)\\
    &\leq \frac{\lambda_e}{\lambda}\left(2+\sum_{\ell=1}^{2f(n)-1}\frac{1}{\ell}\right)\\
    &\leq \frac{\lambda_e}{\lambda}(2+\log{2} + \log{f(n)} + \gamma) \label{final-range1}
\end{align}
where $\gamma \approx 0.577$ is the Euler-Mascheroni constant.

\subsection{Range~2}\label{sec3b}
The upper bound for the recursion for $Y$ is,
\begin{align}
    Y \leq& K\frac{\lambda_e}{\lambda}\left(\frac{1}{\frac{f(n)+1}{n}+\frac{f(n)+1}{2}}\right)\left(\frac{f(n)+1}{2}\right)\notag\\
    &\times\left(1+\sum_{i=f(n)+2}^{n-f(n)}\prod_{j=f(n)+2}^i \frac{1}{1+\frac{2j}{n(f(n)+1)}}\right) \label{eq_Y_rub}
\end{align}
where
\begin{align}
    K &= \prod_{j=1}^{f(n)}\frac{\frac{2jf(n)-j(j-1)}{2f(n)}}{\frac{j+1}{n} + \frac{2(j+1)f(n)-j(j+1)}{2f(n)}}\\
    &\leq \prod_{j=1}^{f(n)}\frac{j}{j+1}\frac{2f(n)-(j-1)}{2f(n)-j}\\
    &= \frac{2}{f(n)+1}
\end{align}
Substituting this in \eqref{eq_Y_rub}, we get,
\begin{align}
    Y \leq& \frac{\lambda_e}{\lambda}\frac{2}{f(n)+1}\frac{2}{f(n)+1}\frac{f(n)+1}{2}\notag\\
    &\times\left(1+\sum_{i=f(n)+2}^{n-f(n)}\prod_{j=f(n)+2}^i \frac{1}{1+\frac{2j}{n(f(n)+1)}}\right)\\
    =& \frac{\lambda_e}{\lambda}\frac{2}{f(n)+1}\left(1+\prod_{j=1}^{f(n)+2} \left(1+\frac{2j}{n(f(n)+1)}\right)\right. \nonumber\\
    &\qquad \qquad \quad \times \left.\sum_{i=f(n)+2}^{n-f(n)}\prod_{j=1}^i \frac{1}{1+\frac{2j}{n(f(n)+1)}}\right)\\
    \leq& \frac{\lambda_e}{\lambda}\frac{2}{f(n)+1}\left(1+\prod_{j=1}^{f(n)+2} \left(1+\frac{2j}{n(f(n)+1)}\right)\right.\nonumber \\
    &\qquad \qquad \qquad \times \left. \sum_{i=1}^{n-f(n)}\prod_{j=1}^i \frac{1}{1+\frac{2j}{n(f(n)+1)}}\right)\label{eq_range_2_2}
\end{align}
Next, we take a logarithm of the $i$th product term in the sum of products term for small enough $i$ and use $\log(1+x)\approx x$, 
\begin{align}
    \!\!\!\!-\log \left(\prod_{j=1}^i \frac{1}{1\!+\!\frac{2j}{n(f(n)+1)}} \right)\!\!=& \sum_{j=1}^i \log\left(1\!+\!\frac{2j}{n(f(n)\!+\!1)}\right) \\
    = & \sum_{j=1}^i \frac{2j}{n(f(n)+1)} \\
    = &\frac{i(i+1)}{n(f(n)+1)}\label{eq29}
\end{align}
In a similar way, we have,
\begin{align}
    \log\left(\prod_{j=1}^{f(n)+2} \left(1+\frac{2j}{n(f(n)+1)}\right) \right)= \frac{f(n)}{n} \label{eq30}
\end{align}
Substituting (\ref{eq29}) and (\ref{eq30}) into \eqref{eq_range_2_2}, we obtain,
\begin{align}
    Y &\leq \frac{\lambda_e}{\lambda}\frac{2}{f(n)+1}\left(1+e^{\frac{f(n)}{n}}\sum_{i=1}^{n-f(n)}e^{-\frac{i(i+1)}{n(f(n)+1)}}\right)\\
    &\leq \frac{\lambda_e}{\lambda}\frac{2}{f(n)+1}\left(1+e^{\frac{f(n)}{n}}\sum_{i=1}^{n}e^{-\frac{i^2}{n(f(n)+1)}}\right)\label{before_riemann}
\end{align}
Now, if $f(n) = o(n)$, then $e^{\frac{f(n)}{n}} \rightarrow 1$, and if $f(n) = \theta(n)$, then $e^{\frac{f(n)}{n}} = C$, where $C$ is a constant. Next, we convert the Riemann sum associated with the summation term in \eqref{before_riemann} into a definite integral, and find its exact value. In order to do so, we use step size $\frac{1}{\sqrt{n(f(n)+1)}}$,
\begin{align}
    \frac{1}{\sqrt{n(f(n)+1)}}\sum_{i=1}^{n}e^{-\frac{i^2}{n(f(n)+1)}} &= \int_0^{\infty} e^{-t^2}dt = \frac{\sqrt{\pi}}{2}
\end{align}
as $n \rightarrow \infty$ and $f(n) = o(n)$, and the step size tending to $0$. On the other hand, if $f(n) = \theta(n)$, then the above integral has lower limit $0$ and upper limit a constant, thus giving,
\begin{align}
    \frac{1}{\sqrt{n(f(n)+1)}}\sum_{i=1}^{n}e^{-\frac{i^2}{n(f(n)+1)}} = L
\end{align}
where $L$ is a constant. Using this, we obtain,
\begin{align}
    \sum_{i=1}^{n}e^{-\frac{i^2}{n(f(n)+1)}} &= L\sqrt{n(f(n)+1)}\\
    &\leq\frac{\sqrt{\pi}}{2}\sqrt{n(f(n)+1)}
\end{align}
when $f(n) = \theta(n)$, and,
\begin{align}
    \sum_{i=1}^{n}e^{-\frac{i^2}{n(f(n)+1)}} &= \frac{\sqrt{\pi}}{2}\sqrt{n(f(n)+1)}
\end{align}
when $f(n) = o(n)$.
Substituting it back in \eqref{before_riemann}, we get,
\begin{align}
    Y &\leq  \frac{\lambda_e}{\lambda}\frac{2}{f(n)+1}\left(1+\frac{\sqrt{\pi}}{2}\sqrt{n(f(n)+1)}\right)\\
    &\approx \sqrt{\pi}\frac{\lambda_e}{\lambda}\frac{\sqrt{n}}{f(n)^{\frac{1}{2}}} \label{final-range2}
\end{align}
\subsection{Range~3}
Following a similar calculation to the calculation of $X$, 
\begin{align}
    Z \leq& 2\frac{\lambda_e}{\lambda} + \frac{\lambda_e}{\lambda}\frac{f(n)+1}{2}\prod_{j=1}^{f(n)}\frac{\frac{2jf(n)-j(j-1)}{f(n)}}{\frac{j+1}{n} + \frac{2(j+1)f(n)-j(j+1)}{f(n)}}\notag\\
    &\times\prod_{j=f(n)+2}^{n-f(n)} \frac{1}{1+\frac{j}{n(f(n)+1)}}\frac{1}{1+\frac{n-f(n)}{n}}\times\Bigg(1+\notag\\
    &\sum_{i=n\!-\!f(n)}^{n-2}\prod_{j=n\!-\!f(n)}^i \frac{\frac{(n-j)f(n)-(n-j)(n-j-1)}{f(n)}}{\frac{j+1}{n}\!+\!\frac{(n-j-1)f(n)-(n-j-1)(n-j-2)}{f(n)}}\Bigg)\label{eq_Z_rub}\\
    \leq& 3\frac{\lambda_e}{\lambda} + \frac{\lambda_e}{\lambda}\notag\\
    &\times\sum_{i=n-f(n)}^{n-2}\prod_{j=n-f(n)}^i \!\!\frac{n-j}{n-j-1}\frac{f(n)-(n-j-1)}{f(n)-(n-j-2)}\!\!\label{eq_Z_rub_after}\\
    \leq& 3\frac{\lambda_e}{\lambda} + \frac{\lambda_e}{\lambda}\sum_{i=n-f(n)}^{n-2} \frac{1}{i}\\
    \leq& \frac{\lambda_e}{\lambda} (3+\log{f(n)}) \label{final-range3}
\end{align}
where we go from \eqref{eq_Z_rub} to \eqref{eq_Z_rub_after} by approximating the product in the first line of \eqref{eq_Z_rub} following the calculation of $K$ in \emph{Range~2}. We drop the product in the second line since it is smaller than $1$, and do the regular upper bound in the third line.

Finally, summing the final terms in the three ranges, i.e., in (\ref{final-range1}), (\ref{final-range2}), (\ref{final-range3}), we obtain the upper bound for the age as,
\begin{align}
    v_1 \leq& X+Y+Z\\
    \leq& \frac{\lambda_e}{\lambda}(5 + \log{2} + 2\log{f(n)} + \gamma)+\sqrt{\pi}\frac{\lambda_e}{\lambda}\frac{\sqrt{n}}{f(n)^{\frac{1}{2}}}\label{eq_ub}
\end{align}
giving the desired result.
\end{Proof}

\begin{table*}[t]
    \centering
    \begin{tabular}{|c|c|c|c|c|c|c|c|c|c|}
    \hline
        $\mathbf{\alpha}$  & 0.1 & 0.2 & 0.3 & 0.4 & 0.5 & 0.6 & 0.7 & 0.8 & 0.9\\
        \hline
        age scaling & $n^{0.45}$ & $n^{0.4}$ & $n^{0.35}$ & $n^{0.3}$ & $n^{0.25}$ & $n^{0.2}$ & $n^{0.15}$ & $n^{0.1}$ & $n^{0.05}$\\ 
        \hline
        $n$ & $0$ & $942$ & $24180$ & $955318$ & $1.22\times 10^8$ & $1.64\times 10^{11}$ & $3.33\times 10^{16}$ & $3.9\times 10^{27}$ & $2.74\times 10^{63}$\\
        \hline
    \end{tabular}
    \vspace{0.2cm}
    \caption{The minimum number of nodes required for the rational functions to dominate the logarithm in the upper bound in \eqref{eq_ub}.}
    \vspace{-0.8cm}
    \label{good_table}
\end{table*}

\section{Special Cases}
\subsection{Fully-Connected Network}
In this case, $f(n) = \frac{n-1}{2}$, hence we have,
\begin{align}
    v_1 \leq \frac{\lambda_e}{\lambda}(2 + \log{(n-1)}) \approx \frac{\lambda_e}{\lambda} \log{n}
\end{align}
which is in accordance with \cite{yates21gossip}.

\subsection{Ring With a Fixed Number of Neighbors}
Suppose each node in the ring network has $2d$ neighbors, where $d$ is a constant, i.e., $f(n)=d$. Then, we have,
\begin{align}
    v_1 \leq \sqrt{\pi}\frac{\lambda_e}{\lambda}\frac{\sqrt{n}}{d^{\frac{1}{2}}}
\end{align}
Hence, in this case, $v_1 = O(\sqrt{n})$. The bi-directional ring falls under this category with $f(n)=d=1$, and we recover \cite{buyukates22ClusterGossip}.

\subsection{Ring With a Rational Number of Neighbors, $f(n)=n^\alpha$}\label{sec_4d}
Here, $f(n)=n^\alpha$, with $0<\alpha<1$. This case covers functions over a vast range between $f(n) = 1$ and $f(n) = \theta(n)$. The version age in this case scales as a rational function,
\begin{align}
    v_1 \leq \sqrt{\pi}\frac{\lambda_e}{\lambda}n^{\frac{1-\alpha}{2}}
\end{align}
Hence, in this case, $v_1=O(n^{\frac{1-\alpha}{2}})$.

\subsection{Ring With $\frac{n}{\log^2{n}} \leq f(n) < n$ Neighbors}
From \cite{yates21gossip}, we know that for a fully-connected network, the version age scales as $\theta(\log{n})$. Since the networks with $f(n)$ considered in this subsection have smaller number of connections, the version age of a single node in these networks is larger. Hence, a lower bound for the version age is $\log{n}$. From (\ref{main_result}), the upper bound is also $\log{n}$. Hence, in this case, the version age of a single node scales as $\theta(\log{n})$.

\section{Observations and Remarks}
\begin{remark}
    Extremal animals have been a topic of study in graph theory for a long time \cite{harary1976extremal}. These are connected subgraphs with minimum or maximum number of neighboring nodes, edges or faces in a graph. In this context, Lemma~\ref{lemma2} finds the minimal edge animal for the general ring network.
\end{remark}
\begin{remark}
    In \cite{srivastava2023age}, it was shown that a two dimensional grid has version age scaling of $O(n^{\frac{1}{3}})$. Each node in the grid network has $4$ neighbors. However, in order to achieve a version age scaling of $O(n^{\frac{1}{3}})$ in the generalized ring network, we need $f(n) = n^{\frac{1}{3}}$, i.e., we need $2n^{\frac{1}{3}}$ neighbors. One way to explain this difference in the requirement for connectivity to achieve the same version age scaling is the following: According to \cite[Remark~5]{srivastava2023age}, we can view the grid as a ring network with $n$ connections which are not local in nature. Hence, although the number of connections in a grid network is far less compared to the generalized ring network, the version age scaling is the same. This shows us that the geometry of a network can affect the version age significantly, and having few connections between nodes far away is better than having relatively dense connections which are local.
\end{remark}

\begin{remark}\label{remark4}
    We say that function $g(n)$ dominates $h(n)$ for a specific value of $n$ if $g(n) \geq 10h(n)$. We consider rational functions $f(n)=n^\alpha$. We want to see the values of $n$ for which the $\sqrt{n}/f(n)^\frac{1}{2}$ term dominates the $\log{f(n)}$ term in the upper bound in \eqref{eq_ub}. In \eqref{eq_ub}, we saw that there are two terms in the upper bound: $\log{f(n)}$ and $\sqrt{n}/f(n)^\frac{1}{2}$. In Section~\ref{sec_4d}, we consider the version age scaling for $f(n)=n^\alpha$, and find that the scaling is $O(n^{\frac{1-\alpha}{2}})$ as $n\rightarrow \infty$. However, as $\alpha$ increases, the rational function grows increasingly slowly and dominates the $\log$ term only at very high values of $n$. We summarize these numbers in Table~\ref{good_table}. We note that up to these values of $n$, the version age is upper bounded by $22\log{n}$, and hence we can consider any $f(n) = \Omega(n^{0.6})$ to have logarithmic scaling in version age for all practical purposes.
\end{remark}

\section{Numerical Results}

We have seen in Section~\ref{sec3}, that the upper bound for version age of the generalized ring network depends on the number of nodes $n$, number of connections $2f(n)$, and the information flow rates $\lambda_e$ and $\lambda$. We choose $\lambda_e = \lambda = 1$ in this section. 

\begin{figure}[t]
    \centering
    \includegraphics[width = \linewidth]{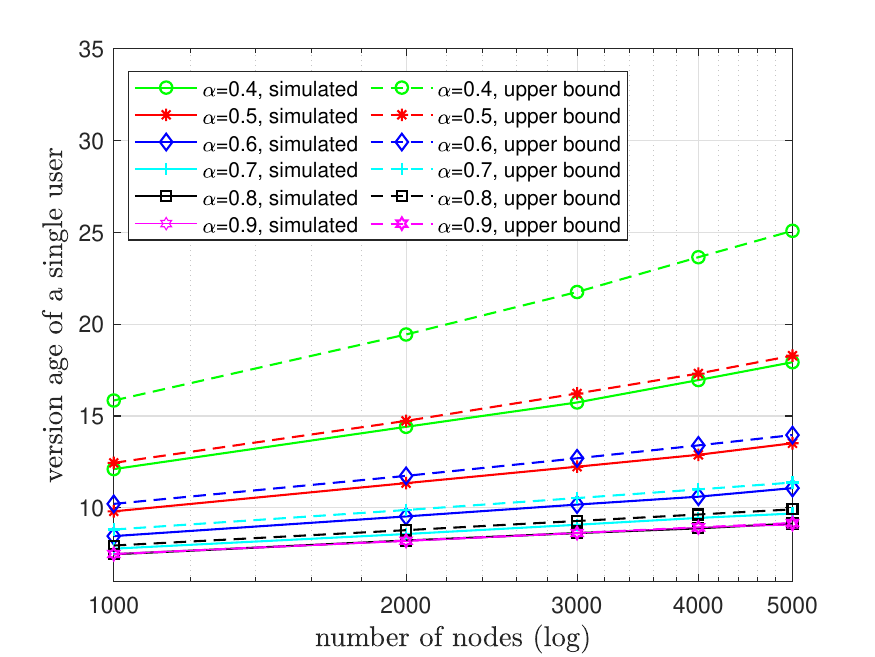}
    \caption{Version age of a single node in the ring network as a function of $\alpha$, where $f(n) = n^{\alpha}$, along with the recursive upper bounds ($n$ in $\log$ scale).}
    \label{fig4}
    \vspace{-0.4cm}
\end{figure}

We plot the variation of the version age for $f(n) = n^\alpha$ for $\alpha=0.4$ to $0.9$. The number of nodes varies from $1000$ to $5000$. Fig.~\ref{fig4} shows that the version age decreases as $\alpha$ increases, which is consistent with our theoretical upper bound result. We also observe that the version age plots in Fig.~\ref{fig4} are straight lines, showing that they have approximate $\log$ scaling for low values of $n$, consistent with Remark~\ref{remark4}.

We have not simulated $\alpha$ between $0$ and $0.3$, because the function $f(n)$ grows slowly. Hence, for small values of $n$, which we are able to run simulations on a PC, the value of $f(n)$ might be constant, even if the number of nodes increases. For $\alpha=0$ to $0.3$, instead of running actual system simulations, we have calculated the upper bound that we obtain from the recursive equations in \eqref{eq_X_rub}, \eqref{eq_Y_rub} and \eqref{eq_Z_rub}, and compared it to the upper bound obtained in \eqref{eq_ub} for $n=10^4$, $10^5$, $10^6$, $10^7$, $10^8$, and observed that the bound gets tight as $\alpha$ increases.

\section{Conclusion}
We considered a gossiping network arranged in a ring. Each node in the network communicates with $f(n)$ nodes on each side, sending and receiving updates. We studied the effect of $f(n)$ on the version age of a node in the network. We found a general upper bound for the version age of a node that depends only on the number of nodes in the network $n$ and $f(n)$. We evaluated the upper bound for several different $f(n)$ regimes.

\bibliographystyle{unsrt}
\bibliography{refs}

\begin{thebibliography}{10}

\bibitem{popovski2022perspective}
P.~Popovski, F.~Chiariotti, K.~Huang, A.~E. Kal{\o}r, M.~Kountouris, N.~Pappas,
  and B.~Soret.
\newblock A perspective on time toward wireless 6{G}.
\newblock {\em Proceedings of the IEEE}, 110(8):1116--1146, August 2022.

\bibitem{kaul2012real}
S.~Kaul, R.~D. Yates, and M.~Gruteser.
\newblock Real-time status: How often should one update?
\newblock In {\em IEEE Infocom}, March 2012.

\bibitem{sun2019age}
Y.~Sun, I.~Kadota, R.~Talak, and E.~Modiano.
\newblock Age of information: A new metric for information freshness.
\newblock {\em Synthesis Lectures on Communication Networks}, 12(2):1--224,
  December 2019.

\bibitem{yatesJSACsurvey}
R.~D. Yates, Y.~Sun, D.~Brown, S.~K. Kaul, E.~Modiano, and S.~Ulukus.
\newblock Age of information: An introduction and survey.
\newblock {\em IEEE Jour. on Selected Areas in Communications},
  39(5):1183--1210, May 2020.

\bibitem{yates2018age}
R.~D. Yates and S.~K. Kaul.
\newblock The age of information: Real-time status updating by multiple
  sources.
\newblock {\em IEEE Transactions on Information Theory}, 65(3):1807--1827,
  March 2018.

\bibitem{banerjee2023re}
S.~Banerjee, S.~Ulukus, and A.~Ephremides.
\newblock To re-transmit or not to re-transmit for freshness.
\newblock May 2023.
\newblock Available online at arXiv:2305.10392.

\bibitem{maatouk20AOII}
A.~Maatouk, S.~Kriouile, M.~Assaad, and A.~Ephremides.
\newblock The age of incorrect information: A new performance metric for status
  updates.
\newblock {\em IEEE/ACM Trans. on Networking}, 28(5):2215--2228, October 2020.

\bibitem{zhong18AoSync}
J.~Zhong, R.~D. Yates, and E.~Soljanin.
\newblock Two freshness metrics for local cache refresh.
\newblock In {\em IEEE ISIT}, June 2018.

\bibitem{cho3BinaryFreshness}
J.~Cho and H.~Garcia-Molina.
\newblock Effective page refresh policies for web crawlers.
\newblock {\em ACM Trans. Database Syst.}, 28(4):390--426, December 2003.

\bibitem{yates21gossip}
R.~D. Yates.
\newblock The age of gossip in networks.
\newblock In {\em IEEE ISIT}, July 2021.

\bibitem{Abolhassani21version}
B.~Abolhassani, J.~Tadrous, A.~Eryilmaz, and E.~Yeh.
\newblock Fresh caching for dynamic content.
\newblock In {\em IEEE Infocom}, May 2021.

\bibitem{melih2020infocom}
M.~Bastopcu and S.~Ulukus.
\newblock Who should {G}oogle {S}cholar update more often?
\newblock In {\em IEEE Infocom}, July 2020.

\bibitem{buyukates22ClusterGossip}
B.~Buyukates, M.~Bastopcu, and S.~Ulukus.
\newblock Version age of information in clustered gossip networks.
\newblock {\em IEEE Jour. on Selected Areas in Information Theory},
  3(1):85--97, March 2022.

\bibitem{kaswan22timestomp}
P.~Kaswan and S.~Ulukus.
\newblock Susceptibility of age of gossip to timestomping.
\newblock In {\em IEEE ITW}, November 2022.

\bibitem{kaswan22jamming}
P.~Kaswan and S.~Ulukus.
\newblock Age of gossip in ring networks in the presence of jamming attacks.
\newblock In {\em Asilomar Conference}, October 2022.

\bibitem{kaswan23nonpoisson}
P.~Kaswan and S.~Ulukus.
\newblock Age of information with non-{P}oisson updates in cache-updating
  networks.
\newblock In {\em IEEE ISIT}, June 2023.

\bibitem{mitra_allerton22}
P.~Mitra and S.~Ulukus.
\newblock {ASUMAN}: Age sense updating multiple access in networks.
\newblock In {\em Allerton Conference}, September 2022.

\bibitem{mitra2023timely}
P.~Mitra and S.~Ulukus.
\newblock Timely opportunistic gossiping in dense networks.
\newblock In {\em IEEE Infocom}, May 2023.

\bibitem{abd2023distribution}
M.~A. Abd-Elmagid and H.~S Dhillon.
\newblock Distribution of the age of gossip in networks.
\newblock {\em Entropy}, 25(2):364--395, January 2023.

\bibitem{srivastava2023age}
A.~Srivastava and S.~Ulukus.
\newblock Age of gossip on a grid.
\newblock In {\em Allerton Conference}, September 2023.
\newblock Also available online at arXiv:2307.08670.

\bibitem{harary1976extremal}
F.~Harary and H.~Harborth.
\newblock Extremal animals.
\newblock {\em Journal of Combinatorics, Information and System Sciences},
  1(1):1--8, January 1976.

\end{thebibliography}

\end{document}